\documentclass{article}
\usepackage{times,fancyhdr}
\usepackage{cite}
\usepackage{graphicx}
\usepackage{axodraw}

\def\beq{\begin{equation}}
\def\eeq{\end{equation}}
\def\beqa{\begin{eqnarray}}
\def\eeqa{\end{eqnarray}}

\title{A comparative study of small $x$ Monte Carlos with and without QCD coherence effects
\footnote{Preprint numbers: LPN11-04, IFT-UAM/CSIC-11-03, FTUAM-11-36}
} 
\author{G. Chachamis$^1$, M. Deak$^2$, A. Sabio Vera$^2$, P. Stephens$^3$ \\ 
\\
$^1$ Paul Scherrer Institut, CH-5232 Villigen PSI, Switzerland\\
\\
$^2$ Instituto de Fisica Teorica UAM/CSIC, Nicolas Cabrera 15, and \\ 
Universidad Autonoma de Madrid, E-28049 Madrid, Spain\\
\\
$^3$ 4800 Oak Grove Dr m/s 138-308, Pasadena 91009-8001, USA}
\begin{document} 

\pagestyle{fancy}
\fancyhead{}
\fancyhead[EC]{G. Chachamis, M. Deak, A. Sabio Vera, P. Stephens}
\fancyhead[EL,OR]{\thepage}
\fancyhead[OC]{A comparative study of small $x$ Monte Carlos with and without QCD coherence effects}
\fancyfoot{} 
\renewcommand\headrulewidth{0.5pt}
\addtolength{\headheight}{2pt} 

\maketitle 

We compare two Monte Carlo implementations of resummation schemes for the description of parton evolution at 
small values of Bjorken $x$. One of them is based on the 
Balitsky-Fadin-Kuraev-Lipatov (BFKL) evolution equation and generates fully 
differential parton distributions in momentum space making use of reggeized gluons. 
The other one is based on the Catani-Ciafaloni-Fiorani-Marchesini (CCFM) partonic kernel where 
QCD coherence effects are introduced. It has been argued that both approaches agree with each 
other in the $x \to 0$ limit. We show that this is not the case for azimuthal angle dependent 
quantities since at high energies the BFKL approach is dominated by its zero conformal spin 
component while the CCFM gluon Green function receives contributions from all conformal spins 
even at very small $x$.

\section{Introduction}

An important challenge in the phenomenology of Quantum Chromodynamics (QCD) 
is to understand what are the dominant effective degrees of freedom underlying the strong 
interaction at high energies. In the limit where the 
center--of--mass energy $\sqrt{s}$ in a collision is much larger than any of the relevant mass scales the  
Balitsky--Fadin--Kuraev--Lipatov (BFKL) 
approach~\cite{BFKL1,BFKL2,BFKL5} appears to be a very useful framework to describe the scattering. In its original formulation this approach is based on the exchange of ``reggeized'' gluons in the  $t$--channel. The interaction among them takes place  
via a gauge invariant (reggeized gluon)-(reggeized gluon)-(gluon) vertex.
This simple effective structure stems from the dominance of the so-called 
multi--Regge kinematics  where gluon cascades are only ordered in longitudinal components.  

The simplicity of the final integral equation 
should not shadow the strong self-consistency of the full BFKL program where tight bootstrap 
conditions linking the reggeization of the gluon with the pomeron wavefunction, dominant in 
diffractive interactions, are fulfilled. The regime of applicability of the leading order BFKL approach 
to phenomenology is limited since it should be valid in a certain window of center-of-mass energies 
and perturbative scales. To extend its range of applicability one should include either higher order 
corrections, going beyond the `multi-Regge" kinematics, include non-linear corrections 
responsible for the restauration of unitarity at very small $x$\footnote{In this work we consider equivalent variables the center-of-mass energy $s$, the DIS Bjorken $x$ and the rapidity $Y$ linking them by $Y \simeq \ln{s} \simeq \ln{1/x}$.} or, as we are going to discuss in this 
paper, to include a more global treatment of collinear regions in phase space using the 
Catani-Ciafaloni-Fiorani-Marchesini (CCFM) equation, which provides a good matching 
from BFKL to the $x \to 1$ regime, at least, as we are going to show in this work, as long as anomalous dimensions and $k_t$-diffusion properties are concerned. 

In this letter we first give a brief introduction to the BFKL and CCFM approaches, 
in section~\ref{BFKLCCFM}. We then describe our Monte Carlo implementations in 
sections~\ref{CCFMMC} and~\ref{BFKLMC}, where we also compare both resummation schemes in 
terms of the detailed exclusive information we can obtain from running the Monte Carlo codes. In 
particular, we show that the UV/IR symmetry in the $k_t$-diffusion present in the BFKL evolution is broken at lower energies in the CCFM equation, to then be slowly restored as $x$ tends to 0. Our 
main result appears when we study the azimuthal angle dependence of the gluon Green 
function in both approaches. We show that the CCFM equation generates a stronger azimuthal 
angle dependence than the BFKL equation. To stress this point we investigate the Fourier 
components of the solution in this angular sector. We show that the projection 
on the zero Fourier component, or zero conformal spin in the BFKL context,  has the highest growth with energy in both schemes. The differences appear in the non-zero Fourier components, or non-zero 
conformal spins, which in the BFKL description fall with energy but in the CCFM solution increase. 
We also discuss the different collinear behaviour of the solutions at different energies. 
Finally we present our conclusions and scope for further investigations. 

\section{Multi-Regge kinematics versus QCD coherence in small $x$ final states}
\label{BFKLCCFM}

Coherence effects are already present in Quantum Electrodynamics where they are responsible for the suppression of soft bremsstrahlung from 
electron--positron pairs. In QCD processes such as $g \rightarrow q {\bar q}$ where a soft gluon is emitted from one of the fermionic lines with an angle larger than the angle of emission in the original $q{\bar q}$ pair probes the total colour charge of this pair. This charge is the same as that of the parent gluon and the final radiation 
takes place as if the soft gluon was emitted from it. This is what we call a colour coherence effect and leads to the angular ordering of sequential gluon emissions. In Deep Inelastic Scattering (DIS) we can focus on the  $(i-1){\rm th}$ emitted gluon with energy $E_{i-1}$ from the proton. A secondary gluon 
radiated from it with a fraction $(1-z_{i})$ of its energy and a transverse momentum $q_{i}$ has the  opening angle 
\begin{eqnarray}
\theta_{i}\approx \frac{q_{i}}{(1-z_{i})E_{i-1}},
\end{eqnarray}
with $z_{i}= E_{i} / E_{i-1}$ and $\left| \theta_i  \right| \ll 1$. Colour coherence leads to angular ordering with increasing opening angles towards the hard scale, in this case the virtuality of the photon $Q^2$, and we have \(\theta_{i+1} > \theta_{i}\), or 
\begin{eqnarray}
\frac{q_{i+1}}{1-z_{i+1}}>\frac{z_{i}q_{i}}{1-z_{i}},
\end{eqnarray}
which in the limit $z_{i},z_{i+1}\ll1$ is equivalent to $q_{i+1}>z_{i}q_{i}$.

In Refs. \cite{Catani:1990yc,Marchesini:1995wr,Ciafaloni:1988ur,Catani:1991gu} 
the BFKL equation for the unintegrated gluon density was written in a form suitable for the study of exclusive quantities:
\begin{eqnarray}
f_{\omega}(\mbox{\boldmath $k$})=f_{\omega}^{0}(\mbox{\boldmath $k$})
+\bar\alpha_{S}\int\frac{d^{2}\mbox{\boldmath $q$}}{\pi q^{2}}\int_{0}^{1}
\frac{dz}{z}z^{\omega}\Delta_{R}(z,k)\Theta(q-\mu)f_{\omega}(\mbox{\boldmath 
$q$}+\mbox{\boldmath $k$}),
\end{eqnarray}
where $\omega$ is the Mellin-conjugate variable to Bjorken $x$, $\mu$ plays the role of a collinear cutoff, \mbox{\boldmath $q$} is the transverse momentum of the emitted gluon, and the reggeized gluon propagator is
\begin{eqnarray}
\Delta_{R}(z_{i},k_{i})=\exp\left(-\bar\alpha_{S}\ln\frac{1}{z_{i}}\ln
\frac{k_{i}^{2}}{\mu^2}\right),
\end{eqnarray}
with \(k_i\equiv|\mbox{\boldmath $k$}_{i}|\), and $\bar\alpha_{S}\equiv \alpha_{S} N_c /\pi$. Solving the equation by iteration real gluon emissions are generated with the corresponding virtual corrections summed to all orders. Since $f_{\omega}$ is inclusive and IR finite it corresponds to a sum over all final states with the $\mu$-dependence cancelling between the real and virtual contributions in the final result.

The DIS structure function is calculated integrating over all $\mu^{2} \leq q_{i}^{2}\leq Q^{2}$ and reads
\begin{eqnarray}
F^{\rm BFKL}_{\omega}(Q,\mu) \equiv \Theta(Q-\mu) + \sum_{r=1}^{\infty}\int_{\mu^{2}}
^{Q^{2}}\prod_{i=1}^{r}\frac{d^{2}\mbox{\boldmath $q$}_{i}}{\pi q_{i}^{2}}
dz_{i}\frac{\bar{\alpha}_{S}}{z_{i}}z_{i}^{\omega}\Delta_{R}(z_{i},k_{i}),
\end{eqnarray}
where $i$ corresponds to the number of gluon emissions, which in the leading order approximation coincides with the number of iterations of the kernel. In Ref.~\cite{Marchesini:1995wr} it was pointed out that coherence effects should  significantly modify each contribution to the final sum with a fixed number of  
emitted gluons, $r$,  whilst preserving the sum. Therefore, care must be taken to account properly for coherence in the calculation of associated distributions. 

Modifying the BFKL formalism to account for 
coherence~\cite{Catani:1990yc,Marchesini:1995wr,Ciafaloni:1988ur,Catani:1991gu}, $F^{\rm BFKL}_\omega$ becomes
\begin{eqnarray}
F^{\rm CCFM}_{\omega}(Q,\mu) &=& \Theta(Q-\mu) \nonumber\\
&&\hspace{-2cm} + \sum_{r=1}^{\infty}\int_{0}^{Q^{2}}
\prod_{i=1}^{r}\frac{d^{2}\mbox{\boldmath $q$}_{i}}{\pi q_{i}^{2}}dz_{i}
\frac{\bar{\alpha}_{S}}{z_{i}}z_{i}^{\omega}\Delta(z_{i},q_{i},k_{i})
\Theta(q_{i}-z_{i-1}q_{i-1}),
\end{eqnarray}
where $\Delta(z_{i},q_{i},k_{i})$ is not a reggeized gluon propagator anymore but stills plays the role of a no-emission factor and takes the CCFM form
\begin{eqnarray}
\Delta(z_{i},q_{i},k_{i})=\exp\left[-\bar\alpha_{S}\ln\frac{1}{z_{i}}\ln
\frac{k_{i}^{2}}{z_{i}q_{i}^{2}}\right];~ ~ k_{i} > q_{i}.
\label{constraint}
\end{eqnarray}
For the first emission $q_{0}z_{0} = \mu$ is chosen. This collinear cutoff is needed only in the first emitted gluon because subsequent collinear 
emissions are already regulated by the angular ordering constraint. 

Both approaches were compared when the rates for emission of a fixed number of resolved gluons, with a transverse momentum larger 
than a given resolution scale $\mu_{R}$, together with any number 
of unresolved ones, was performed in Ref.~\cite{Forshaw:1998uq}.
The $\mu_{R}$ scale is constrained by the collinear cutoff and the hard scale, $\mu \ll \mu_{R}\ll Q$. It was found that jet rates in both multi-Regge (BFKL) and coherent (CCFM) schemes are the same. 
When coherence is introduced the singularities at 
$\omega = 0$ are stronger than in the BFKL approach but the extra logarithms cancel in the sum of all the contributions to the jet rates. Using a generating function for the jet multiplicity it is possible to prove that this is true to all orders in the coupling~\cite{Webber:1998we,Forshaw:1999yh}. The same generating function is obtained for BFKL and CCFM, with the $p$-th central moment of the jet multiplicity distribution being a polynomial in $\frac{{\bar \alpha}_s}{\omega} \ln{\frac{Q}{\mu_R}}$ of degree $2p-1$, showing that the distribution becomes relatively narrow in
the limit of very small $x$ and large $Q/\mu_{R}$~\cite{Webber:1998we}.

In Refs.~\cite{Ewerz:1999fn} the subject was developed further and subleading logarithms were included to calculate the minijet multiplicity associated to Higgs production at hadron colliders. In~\cite{Salam:1999ft} it was also shown that for any sufficiently inclusive observable the CCFM formalism should lead to the same results as the BFKL equation. For the interested reader good reviews devoted to the implementation of CCFM in Monte Carlo event generators can be found in, {\it e.g.},~\cite{Dittmar:2005ed,Alekhin:2005dx,Andersen:2006pg,Jung:2010si}. 
For Monte Carlo implementations of the BFKL approach see Refs.~\cite{Schmidt:1996fg,Andersen:2003an,Dobbs:2004bu,SabioVera:2003xi, Alekhin:2005dx,Dittmar:2005ed,Andersen:2006pg,SabioVera:2006rp,Jung:2008tq}. More recent 
results related to the implementation of models with unitarization in the CCFM 
formalism can be found in Refs.~\cite{Avsar:2009pf,Avsar:2010ia}

\section{Monte Carlo implementation of the CCFM evolution and numerical results}
\label{CCFMMC}

The numerical implementation of the CCFM equation we use in the present 
work is the Monte Carlo event generator {\large S}MALLX 
developed by Marchesini and Webber in Ref.~\cite{Marchesini:1990zy}. In this code the CCFM gluon Green function, interpreted as the unintegrated gluon structure function 
$f^{\rm CCFM} \left({\bf k}_a,{\bf k}_b,x\right)$ when 
acting on the initial condition shown below, can be written in the iterative form
\begin{eqnarray}
f^{\rm CCFM} \left({\bf k}_a,{\bf k}_b,x\right) &=& \delta \left(x-x_0\right) 
\delta^{(2)} \left({\bf k}_a - {\bf k}_b\right) \Delta_S \left(\eta, q_0\right)
\theta \left(\eta- q_0\right)\nonumber\\
&&\hspace{-3cm}+\sum_{n=1}^\infty \int \left[\prod_{i=1}^n d {\rm PS}_i\right]
\theta \left(\eta - z_n l_n\right) \Delta_S \left(\eta, z_n l_n\right)
\delta \left(x-x_n\right) \delta^{(2)}\left({\bf k}_a - {\bf k}_n\right).
\label{CCFMiterative}
\end{eqnarray}
As we mentioned in the previous section, $q_0$ is the collinear cutoff for the first real emission. ${\bf k}_i$ corresponds to the transverse momentum of the exchanged gluons, with $x_i$ being their longitudinal momentum fraction. $l_i = q_i 
/ (1-z_i)$ is the rescaled transverse momentum of the emitted gluons. $\eta$ is 
the upper limit of the phase space integration. The Sudakov form factor reads
\begin{eqnarray}
\Delta_S \left(l_i,z_{i-1} l_{i-1}\right) &=& {\rm exp}
\left[-2 {\bar \alpha}_s 
\int_{z_{i-1} l_{i-1}}^{l_{i}} \frac{dl}{l} \int_0^{1-\frac{\lambda}{l}}\frac{dz}{1-z}\right],
\end{eqnarray}
where $\lambda$ is a cutoff for soft singularities. The final result is independent of 
$\lambda$ in the limit $\lambda \to 0$. The probability of emission of each of the 
real gluons can be written in the form
\begin{eqnarray}
d {\rm PS}_i = \Delta_S \left(l_i,z_{i-1} l_{i-1}\right)  P \left(z_i,q_i,k_i\right) 
\theta \left(l_i - z_{i-1} l_{i-1} \right) \theta \left(1-z_i- \frac{\lambda}{l_i}\right)
\frac{d^2 {\bf l}_i}{\pi l_i^2} d z_i,
\end{eqnarray}
with the gluon splitting function being
\begin{eqnarray}
P \left(z_i,q_i,k_i\right)  &=& {\bar \alpha}_s \left(\frac{1}{1-z_i}+\frac{\Delta(z_i,q_i,k_i)}{z_i}\right).
\end{eqnarray}

Our target is to compare this numerical implementation of the solution to the 
CCFM equation with the corresponding one in the BFKL approach. For this we 
have studied several distinctive distributions which will allow for the comparison. 
In this letter we focus on the leading order approximation in both approaches 
keeping the coupling fixed. We will investigate higher order corrections in future 
publications. 

\begin{figure}[htbp]
  \centering
  \includegraphics[width=10cm]{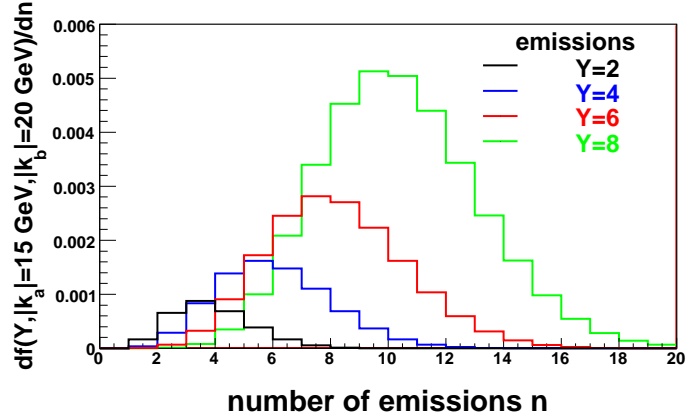}
  \caption{Distribution in the contributions to the CCFM gluon Green function with a fixed number of emitted gluons, plotted for different values of the center-of-mass energy.}
  \label{CCFMnumber}
\end{figure}

\begin{figure}[htbp]
  \centering
  \includegraphics[width=8cm]{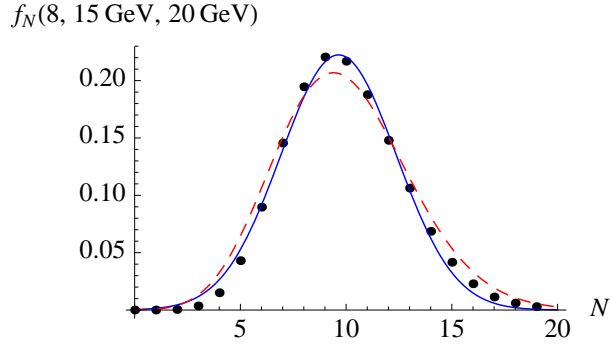}
  \caption{Gaussian (straight) and Poissonian (dashed) fits to the distribution in the contributions to the CCFM gluon Green function with a fixed number of emitted gluons, plotted for rapidity $Y=8$ and   fixed values of the external momenta.}
  \label{GaussianvsPoisson}
\end{figure}

If we represent the CCFM solution in the iterative form of Eq.~(\ref{CCFMiterative}) 
we can test its convergence investigating what is the contribution to 
$f^{\rm CCFM}$ stemming from a fixed number of real gluon emissions. For 
a given $x=e^{-Y}$ when $Y$ is larger configurations with more emissions have 
a bigger weight. This is seen in Fig.~\ref{CCFMnumber} where we also notice that 
the maximum of the distribution, with ${\bar \alpha}_s=0.2$, for $Y = 2, 4, 6, 8$ lies, 
respectively, at $n=4,5,7,9$.  
The chosen external momentum scales are $k_a=15, k_b=20$ GeV. In 
Fig.~\ref{GaussianvsPoisson} we observe that the shape of the distribution is 
better reproduced by a Gaussian than by a Poissonian fit. 

\begin{figure}[htbp]
  \centering
  \includegraphics[width=11cm]{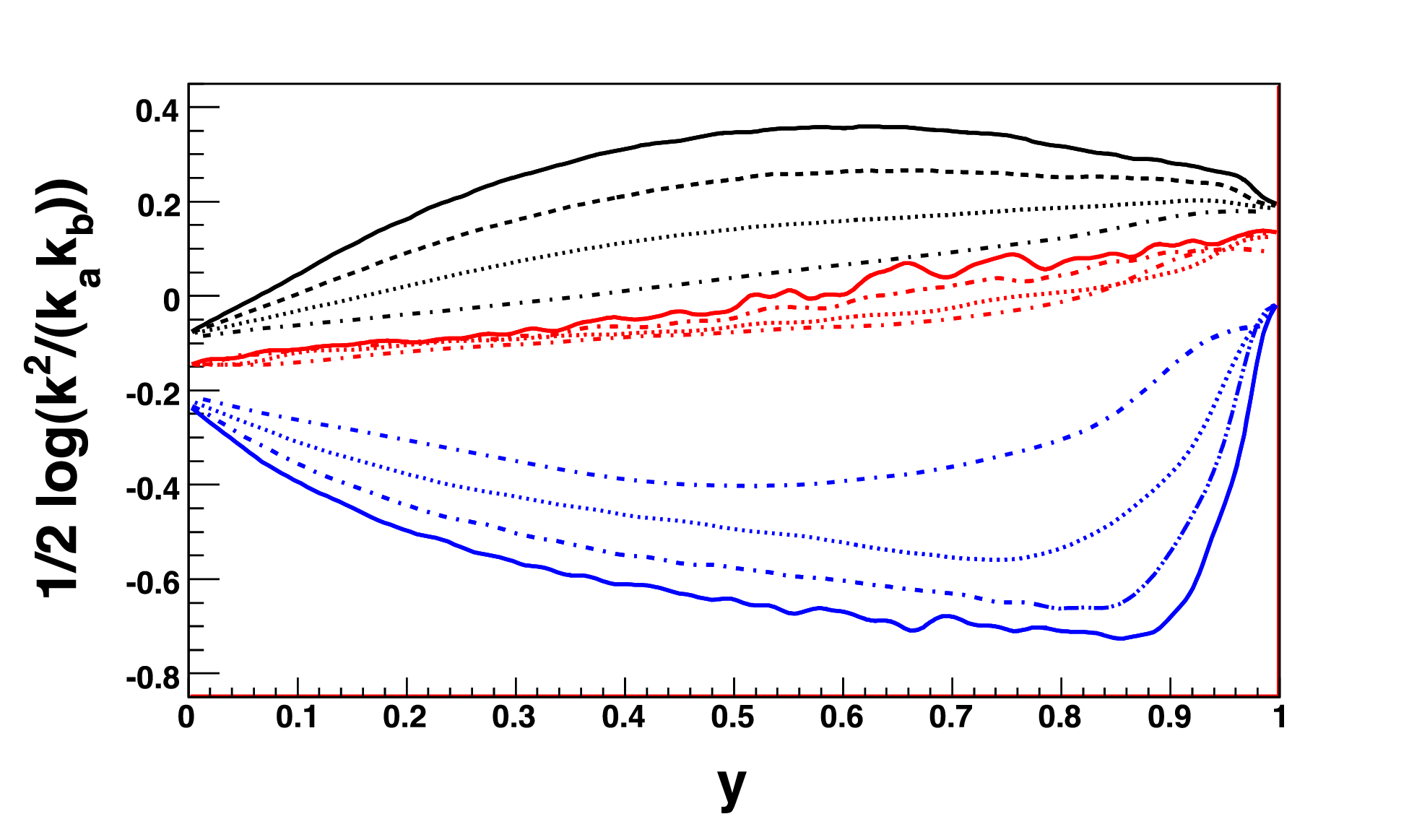}
  \caption{Distribution of transverse scales in the evolution with $x$ of the CCFM equation.}
  \label{CCFMdiffusion}
\end{figure}

As the gluon emissions take place it is more likely to populate regions of phase 
space far away from the transverse scales present in the initial condition. We have studied 
this point in Fig.~\ref{CCFMdiffusion} where we show the mean deviation from the central value of the 
typical transverse momentum of the emitted gluons from the external scales 
$k_a, k_b$. We continue choosing ${\bar \alpha}_s = 0.2$ and we now take 
$k_a=15$ GeV and $k_b=20$ GeV. The central lines show the maximum of the distribution in the momentum of the emitted gluon $k$ for a given rapidity $Y$, showing all plots with different total rapidities in a single figure normalizing them to the interval from zero to one. The two outermost lines correspond to $Y=8$ with the total $Y$ decreasing as we move closer to the central line. 
It is very interesting to see how the diffusion into the infrared region, where 
$k^2 < k_a k_b$, dominates the evolution. To quantify this statement we have 
calculated the ratio 
\begin{eqnarray}
{\cal R}_{\rm UV/IR} &=& \frac{\rm Area_{~UV}}{\rm Area_{IR}}
\label{UVIRRatio}
\end{eqnarray}
of the area above (ultraviolet diffusion) over the area below (infrared diffusion) 
the central line. For $Y = 2, 4, 6 ,8$ we obtain ${\cal R}_{\rm UV/IR} = 0.42, 0.43, 0.44, 0.47$. From these numbers we can see that the contribution of the infrared 
region is enhanced as we decrease the available energy in the scattering process. If we introduce a cut in the number of gluons considering 
only those contributions to the gluon Green function with more than 10 emissions then 
we obtain ${\cal R}_{\rm UV/IR} = 0.39, 0.48, 0.51$ for, respectively, 
$Y= 4, 6, 8$. This indicates that high-multiplicity contributions are more dominated by infrared effects for low total rapidities than the 
low-multiplicity configurations, while converging to a more UV/IR symmetric structure for larger $Y$. This is natural since they dominate the full Green function in this region. 

\begin{figure}[htbp]
  \centering
  \includegraphics[width=10cm]{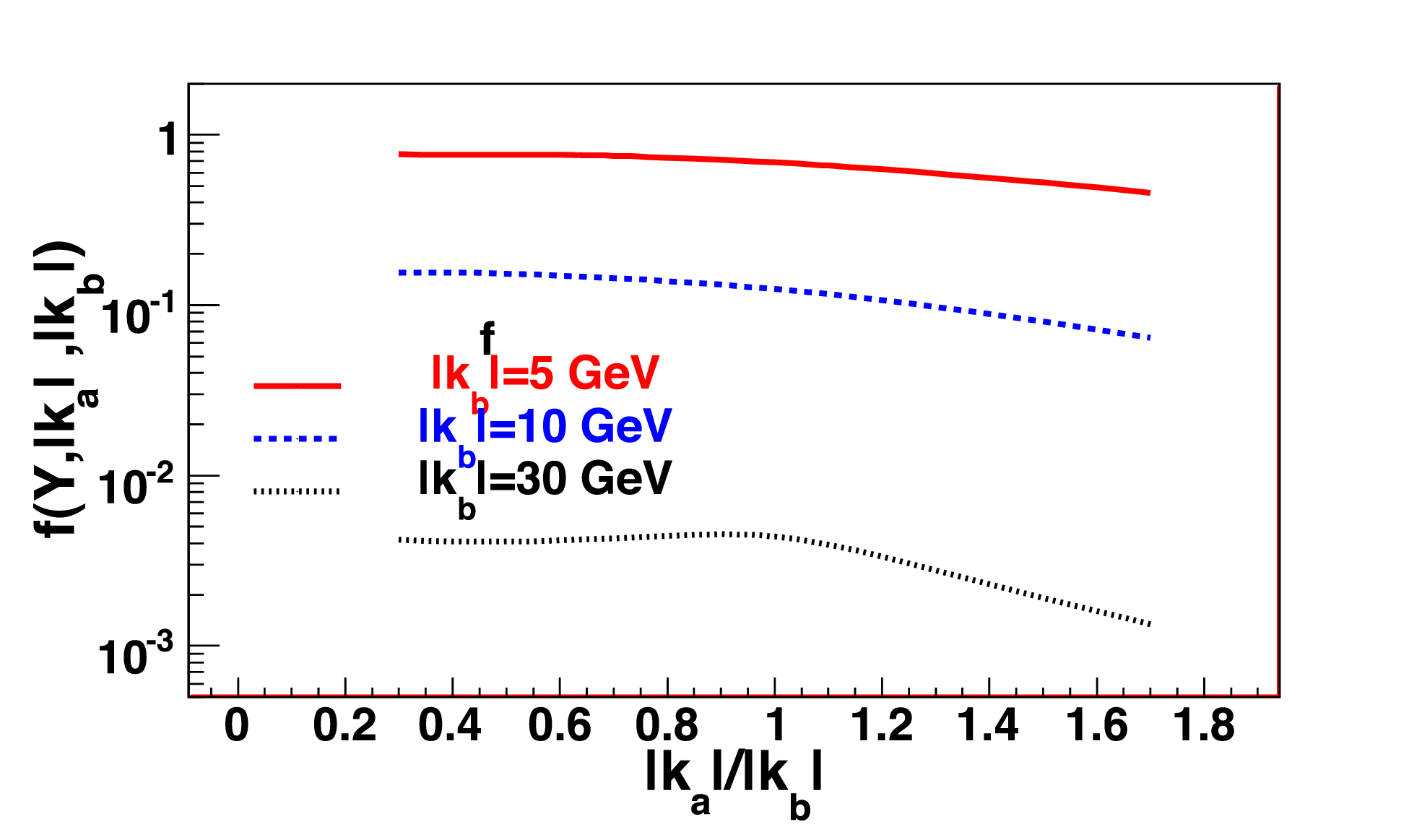}
  \caption{Collinear behaviour of the CCFM gluon Green function for different values of the reference scale, $k_b$.}
  \label{CCFMcollinear}
\end{figure}

\begin{figure}[htbp]
  \centering
  \includegraphics[width=10cm]{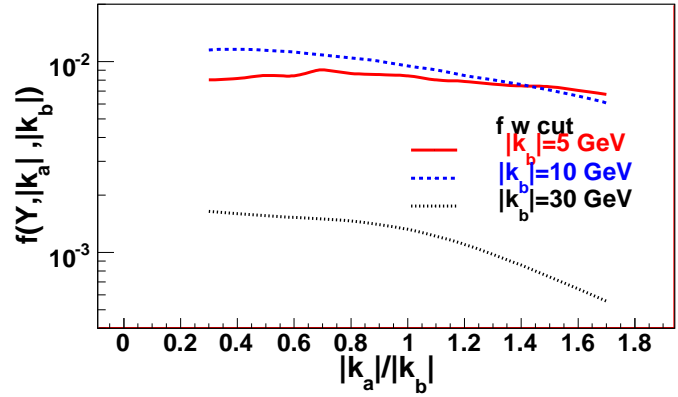}
  \caption{Collinear behaviour of the CCFM gluon Green function for different values of the reference scale, $k_b$, considering the contributions of configurations with more than 10 emissions.}
  \label{CCFMcollinearcut}
\end{figure}

When comparing with the BFKL results, it is interesting to investigate the collinear/anticollinear behaviour 
of the solution to the CCFM equation. This can be done by studying, with a fixed rapidity $Y$, the regions 
with large/small ratio of the external scales $k_b/k_a$. We have done this in Fig.~\ref{CCFMcollinear} for 
different values of the `reference scale" $k_b = 5, 10, 30$ GeV.  
The main conclusion of this analysis is that at lower values of the external scale $k_b$ the gluon Green 
function becomes much flatter as a function of the variation of the other external scale $k_a$. One is 
naturally tempted to relate this behaviour to a possible approximate ``conformal invariance" present in this 
limit. If we now introduce a cut in the number of emissions and keep only those contributions with more than ten gluons in the final state then we obtain Fig.~\ref{CCFMcollinearcut}. We can see that the approximate scale invariance present in the low $k_b$ cases 
remains in the low-multiplicity configurations, being a quite universal feature of the CCFM radiation.

\begin{figure}[htbp]
  \centering
  \includegraphics[width=10cm]{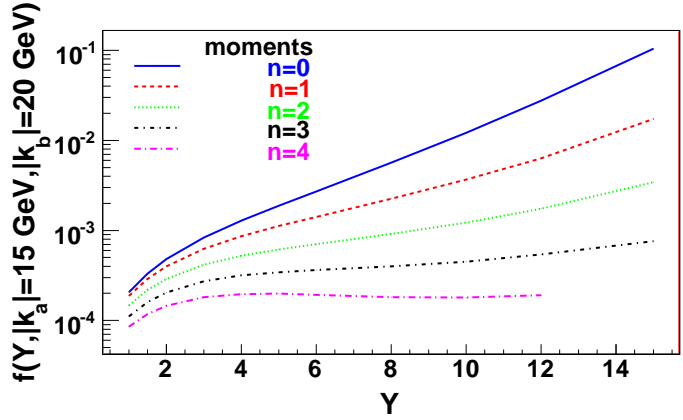}
  \caption{Variation with rapidity of the different components of the Fourier expansion on the azimuthal angle of the CCFM gluon Green function.}
  \label{CCFMFourier}
\end{figure}

\begin{figure}[htbp]
  \centering
  \includegraphics[width=10cm]{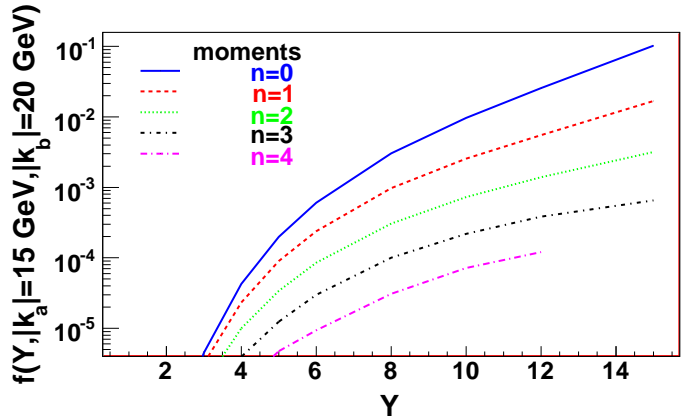}
  \caption{Variation with rapidity of the different components of the Fourier expansion on the azimuthal angle of the CCFM gluon Green function,  considering the contributions of configurations with more than 10 emissions.}
  \label{CCFMFouriercut}
\end{figure}

We now focus on the azimuthal angle dependence of our numerical solution. The azimuthal angle, 
$\theta$, we are interested in is that formed by the two two-dimensional external vectors 
${\bf k}_a$ and ${\bf k}_b$. For this we extract the corresponding Fourier components using
\begin{eqnarray}
f^{\rm CCFM}_n \left(k_a, k_b, x\right) &=& 
\int_0^{2 \pi} \frac{d\theta}{2 \pi} \, f^{\rm CCFM} \left({\bf k}_a, {\bf k}_b, x\right) 
\cos{\left(n ~ \theta\right)}.
\end{eqnarray}
In Fig.~\ref{CCFMFourier} we observe that the $n=0$ component is the dominant one, with the $n>0$ 
components also growing with energy but at a slower pace. We will see that this is completely different to 
the BFKL case. If we impose the high-multiplicity cut then we can see in Fig.~\ref{CCFMFouriercut} that at 
larger $Y$ the Green function is completely dominated by the many-gluon configurations. We can also 
conclude that the growth of all Fourier components is a common feature at any multiplicity. 

\begin{figure}[htbp]
  \centering
  \includegraphics[width=10cm]{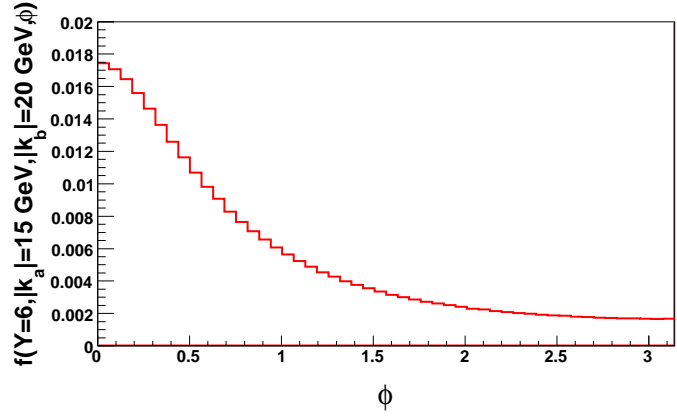}
  \caption{Full dependence on the azimuthal angle of the CCFM gluon Green function.}
  \label{CCFMtheta}
\end{figure}

\begin{figure}[htbp]
  \centering
  \includegraphics[width=10cm]{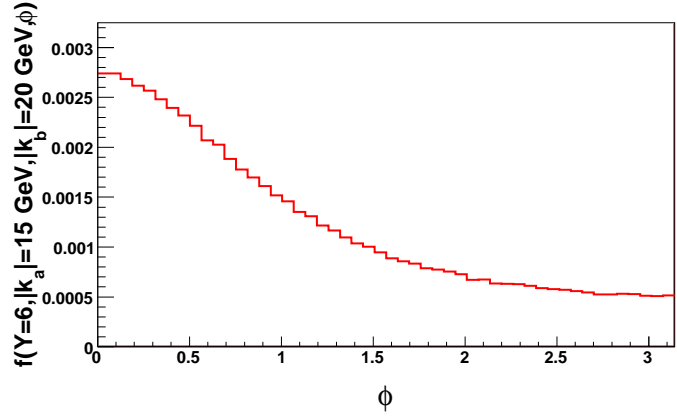}
  \caption{Full dependence on the azimuthal angle of the CCFM gluon Green function, considering the contributions of configurations with more than 10 emissions.}
  \label{CCFMthetacut}
\end{figure}

\begin{figure}[htbp]
  \centering
  \includegraphics[width=10cm]{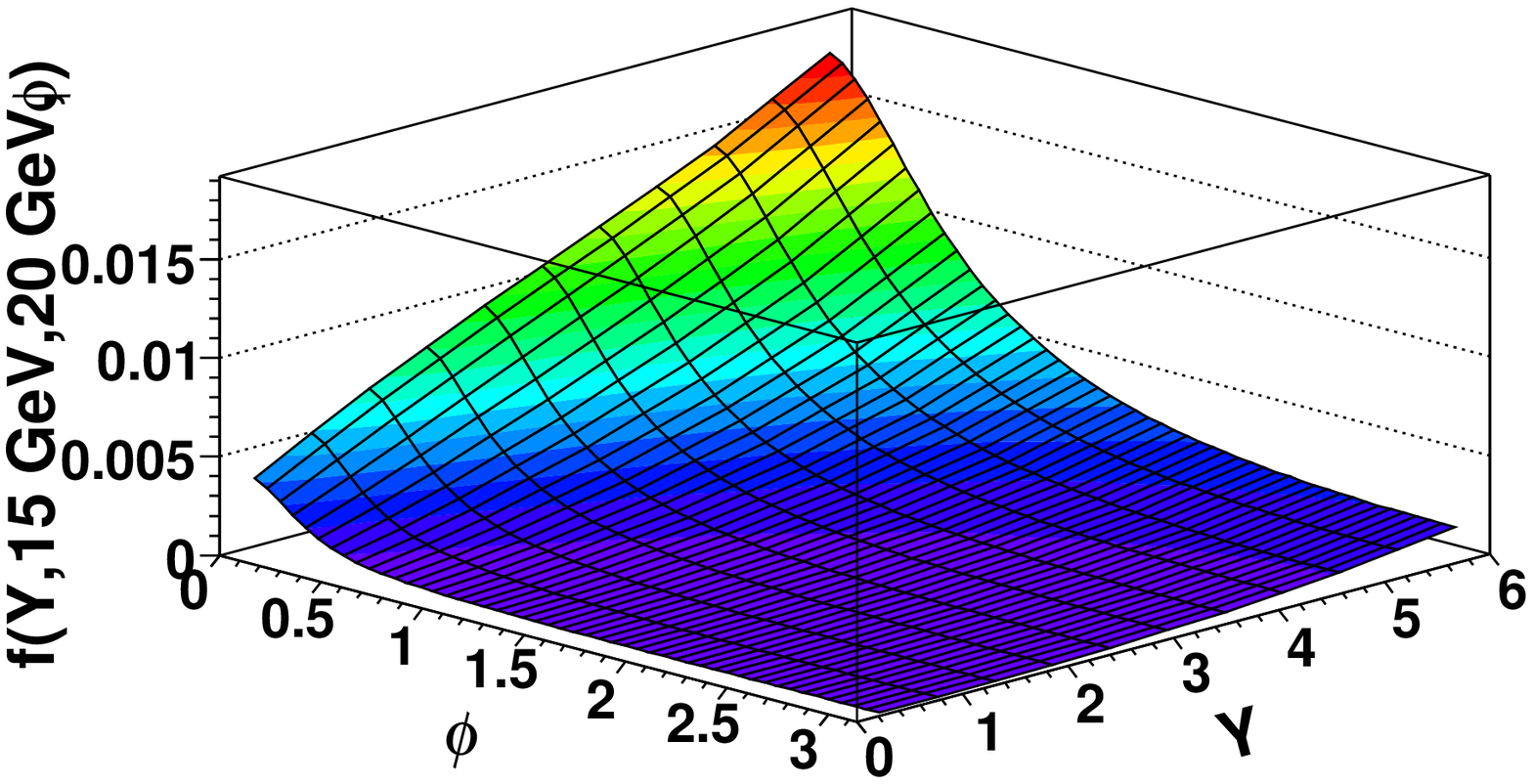}
  \caption{The combined dependence of the CCFM gluon Green funtion on the azimuthal angle and rapidity, for fixed values of the modulus of the boundary momenta.}
  \label{CCFMridge}
\end{figure}

\begin{figure}[htbp]
  \centering
  \includegraphics[width=10cm]{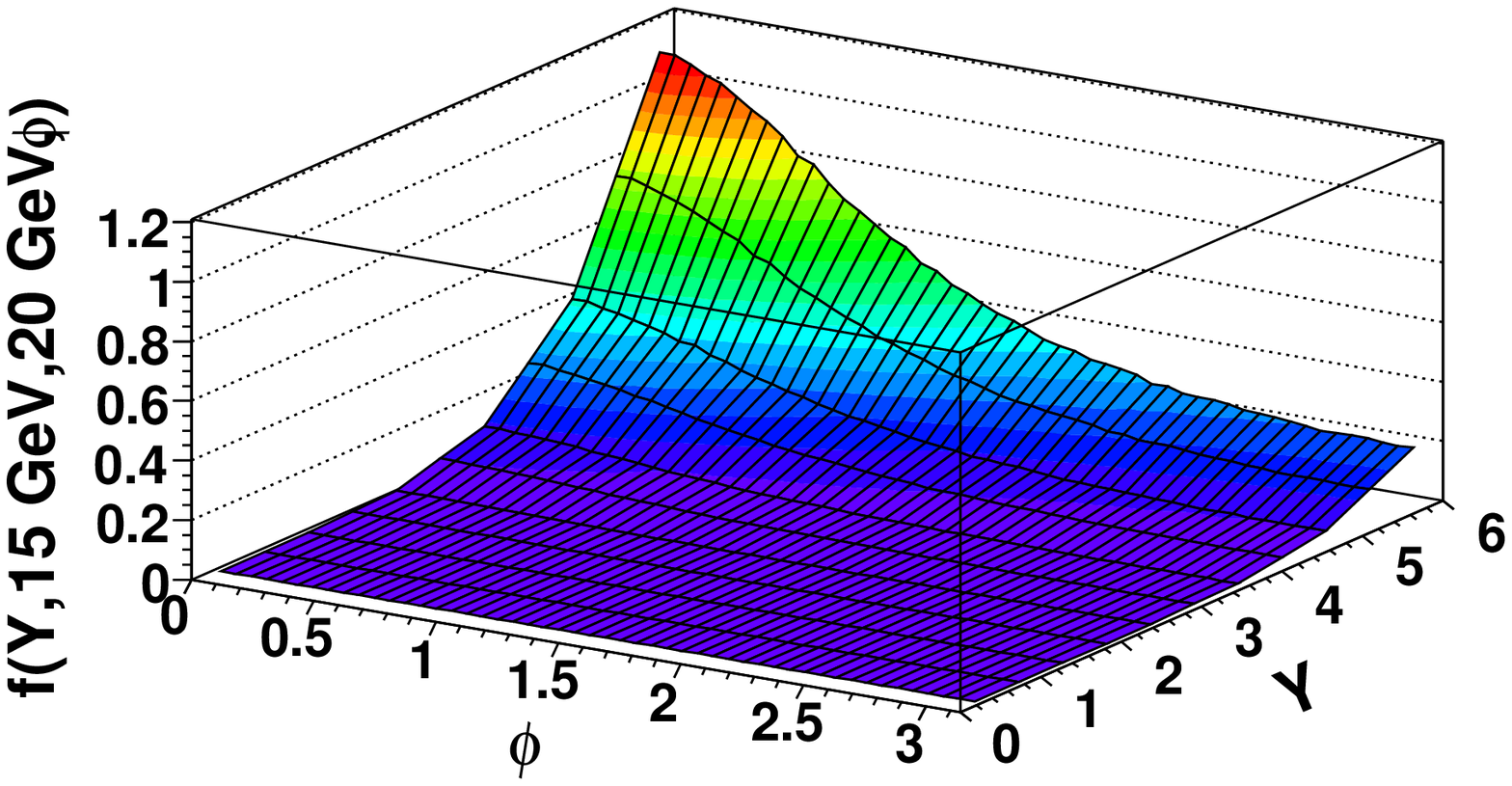}
  \caption{The combined dependence of the CCFM gluon Green funtion on the azimuthal angle and rapidity, for fixed values of the modulus of the boundary momenta, considering the contributions of configurations with more than 10 emissions.}
  \label{CCFMridgecut}
\end{figure}

To wrap up this section we consider the full $\theta$ dependence in Fig.~\ref{CCFMtheta}. It is clear that 
the bulk of the configurations live in the region with ${\bf k}_a$ ``back-to-back" 
with ${\bf k}_b$ 
($\theta = 0$)  with a Gaussian-like spread towards other angular settings. 
At high multiplicities the ${\bf k}_a$ and ${\bf k}_b$ momenta tend to be more decorrelated in the azimuthal 
angle as seen in Fig.~\ref{CCFMthetacut}.
The combined $(\theta, Y)$ dependence is shown in Fig.~\ref{CCFMridge}.
When only the high-multiplicity configurations are kept we obtain a slower variation in angles in 
Fig.~\ref{CCFMridgecut}.

\section{Monte Carlo implementation of the BFKL evolution and numerical results}
\label{BFKLMC}

Our numerical solution of the BFKL equation is based on the 
following iterative representation for the LO gluon Green function:
\begin{eqnarray}
f^{\rm BFKL} \left({\bf k}_a, {\bf k}_b,{\rm Y}\right) &=& 
e^{2 \,\omega_0 \left(- {k}_a^2\right){\rm Y}} 
\Bigg\{\delta^{(2)}\left({\bf k}_a- {\bf k}_b\right)\nonumber\\
&&\hspace{-3cm}+ \sum_{n=1}^\infty \prod_{i=1}^n {\bar \alpha}_s 
\int \frac{d^2{{\bf k}_i}}{\pi {k}_i^2} 
\theta\left({k}_i^2-\lambda^2\right)
\int_0^{y_{i-1}} \hspace{-0.5cm}dy_i \, e^{2 \, \omega^{(i,i-1)}_0 y_i} 
\delta^{(2)}\left({\bf k}_a- {\bf k}_b+\sum_{l=1}^n {\bf k}_l\right)
\hspace{-0.2cm}\Bigg\},
\label{BFKLsolution}
\end{eqnarray}
where
\begin{eqnarray}
\omega^{(i,i-1)}_0 &\equiv&  \omega_0 
\left(- \left({\bf k}_1+\sum_{l=1}^i {\bf k}_l\right)^2\right)
- \omega_0 
\left(- \left({\bf k}_1+\sum_{l=1}^{i-1} {\bf k}_l\right)^2\right).
\end{eqnarray}
Here we have used this notation for the gluon Regge trajectory:
\begin{eqnarray}
\omega_0 (t) &=& -\frac{{\bar \alpha}_s \, {q}^{{2}}}{4 \pi} 
\int \frac{d^2 {\bf k}}{{k}^{2}({\bf q}-{\bf k})^2} ~\simeq~  
-\frac{{\bar \alpha_s}}{2} \log{\frac{{q}^2}{\lambda^2}},
\label{trajectory}
\end{eqnarray} 
where $t = - {q}^{2}$ and 
$\lambda$ is an infrared regulator. The final result is independent of 
$\lambda$ in the limit of small $\lambda$. The normalization of our numerical implementation corresponds to the following analytic form of the Green function:
\begin{eqnarray}
f^{\rm BFKL} \left({\bf k}_a, {\bf k}_b,{\rm Y}\right)  =\frac{1}{\pi k_a k_b} 
\sum_{n=-\infty}^{\infty} \int \frac{d\omega }{2 \pi i}\, e^{\omega Y}
\int \frac{d \gamma}{2 \pi i} 
\left(\frac{k_a^2}{k_b^2}\right)^{\gamma-\frac{1}{2}}
\frac{e^{i n \theta}}{\omega - \bar{\alpha}_s \chi(n,\gamma)},
\end{eqnarray}
where $\theta$ is the azimuthal angle between the ${\bf k}_a$ and ${\bf k}_b$ 
transverse momenta. 
In the case of elastic scattering $n$ can be interpreted as a conformal spin in the unitary principal series representation of 
$SL(2,C)$~\cite{Lipatov:1985uk}. 
The eigenvalue of the BFKL kernel is 
\begin{eqnarray}
\chi(n,\gamma) &=& 2 \Psi(1) - \Psi\left(\gamma+\frac{n}{2}\right)-\Psi\left(1-\gamma+\frac{n}{2}\right).
\end{eqnarray}

\begin{figure}[htbp]
  \centering
  \includegraphics[width=10cm,angle=0]{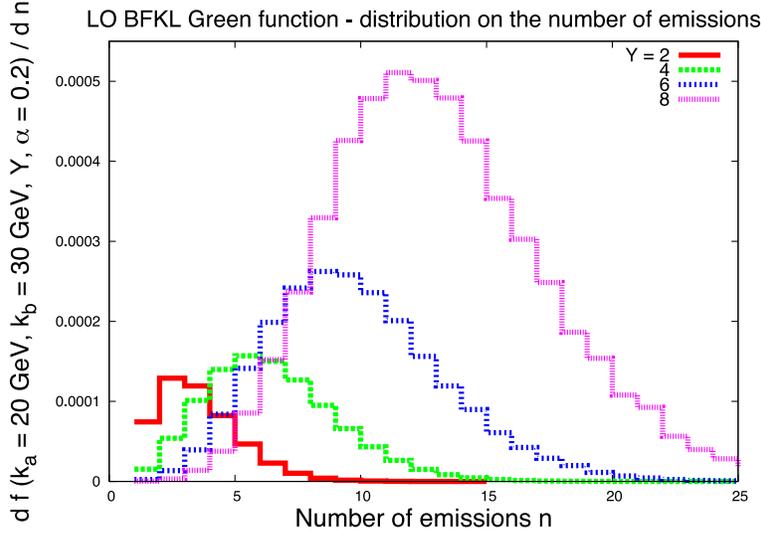}
    \caption{Distribution in the contributions to the BFKL gluon Green function with a fixed 
    number of emitted gluons, plotted for different values of the center-of-mass energy.}
  \label{BFKLemissions}
\end{figure}

The Monte Carlo analysis is useful because it allows for a more detailed study of different 
distributions. In particular, we can, as we did in the CCFM case, investigate the convergence 
of the iterative solution in terms of the contributions with a fixed number of emissions for 
a given value of $Y \simeq \log{1/x}$. We have shown this distribution in 
Fig.~\ref{BFKLemissions} for $Y=2,4,6,8$, $k_a=15$ GeV and $k_b=20$ GeV and 
${\bar \alpha}_s = 0.2$. We observe a very similar pattern to the one present in the CCFM case: a broadening 
towards larger number of emissions for large rapidities with the maxima being at 
$n = 2, 5, 8, 11$ for, respectively, $Y=2,4,6,8$. The main difference here is that in 
the BFKL case the distribution is clearly Poissonian. 

The diffusion picture in BFKL at leading order is well understood. We can average over the azimuthal angle:
\begin{eqnarray}
{\bar f}^{\rm BFKL}  \left(k_a,k_b,{\rm Y}\right) &\equiv&  
\frac{1}{2 \pi} \int_0^{2 \pi} f^{\rm BFKL} \left({\bf k}_a, {\bf k}_b,{\rm Y}\right)
\nonumber\\
&=&\frac{1}{\pi k_a k_b}
\int \frac{d\omega}{2 \pi i} \int \frac{d\gamma}{2 \pi i} 
\left(\frac{k_a^2}{k_b^2}\right)^{\gamma-\frac{1}{2}} 
\frac{e^{\omega {\rm Y}}}{\omega - {\bar \alpha}_s \chi (\gamma)},
\end{eqnarray} 
where  $\chi(\gamma)$ has poles at integer values of $\gamma$ with the 
physical region corresponding to $0<\gamma<1$. At asymptotic values of Y the 
dominant region of integration corresponds to $\gamma \simeq \frac{1}{2}$. 
The $\gamma \leftrightarrow 1- \gamma$ symmetry of $\chi$ indicates that there exists a 
symmetric diffusion into infrared and ultraviolet modes: since the pole at 
$\gamma \simeq 0 \,(1)$ corresponds to virtualities in the internal 
propagators smaller (larger) than the external scales, it leads to infrared 
(ultraviolet) diffusion. To be more precise: for 
very large values of ${\bar \alpha}_s {\rm Y}$ we can write 
\begin{eqnarray}
{\bar f}^{\rm BFKL}  \left(k_a,k_b,{\rm Y}\right)
&=&\frac{1}{\pi k_a k_b} \int \frac{d\gamma}{2 \pi i} 
\left(\frac{k_a^2}{k_b^2}\right)^{\gamma-\frac{1}{2}} 
e^{\chi (\gamma){\bar \alpha}_s {\rm Y}},
\label{angavephi}
\end{eqnarray}
and evaluate the integral using the saddle point approximation around the 
minimum of $\chi$:
\begin{eqnarray}
\chi (\gamma) &\simeq& 4 \log{2} + 14 \, \zeta_3 \left(\gamma-\frac{1}{2}\right)^2 
+ \cdots
\label{symmgamma}
\end{eqnarray}
to obtain the expression
\begin{eqnarray}
{\bar f}^{\rm BFKL}  \left(k_a,k_b,{\rm Y}\right)
&\simeq&\frac{1}{2 \pi k_a k_b} e^{ {\bar \alpha}_s 4 \log{2} {\rm Y}} 
\frac{1}{\sqrt{ 14 \pi \zeta_3{\bar \alpha}_s {\rm Y}}} 
e^{\frac{-t^2}{56 \zeta_3{\bar \alpha}_s {\rm Y}}},
\end{eqnarray}
with $t \equiv \log{(k_a^2 / k_b^2)}$. It is easy to verify that the function
\begin{eqnarray}
\Phi \left( k_a, k_b, {\rm Y} \right) &\equiv& k_a \, k_b \,
{\bar f}^{\rm BFKL}  \left(k_a,k_b,{\rm Y}\right)
\end{eqnarray}
asymptotically fulfills the diffusion equation
\begin{eqnarray}
\frac{\partial \Phi}{\partial ({\bar \alpha}_s {\rm Y})} &=& 
4 \, \log{2} \, \Phi + 14 \, \zeta_3 \,\frac{\partial^2 \Phi}{\partial t^2},
\end{eqnarray}
which is independent of the sign of $t$ and therefore, going back to the definition in 
Eq.~(\ref{UVIRRatio}), ${\cal R}_{\rm UV/IR}=1$  in the BFKL case. It is interesting to note that this diffusive behaviour is 
only driven by the anomalous dimension $\gamma$ while the $n \neq 0$ do not play any role. The fact that in the CCFM case 
the ratio ${\cal R}_{\rm UV/IR}$ tends to 1 for very large rapidities is in 
agreement with the fact in the small $x$ limit the CCFM and 
BFKL approaches have the same asymptotic limit for the anomalous dimensions governing the scale variation of DIS structure functions. 

\begin{figure}[htbp]
  \centering
  \includegraphics[width=10cm,angle=0]{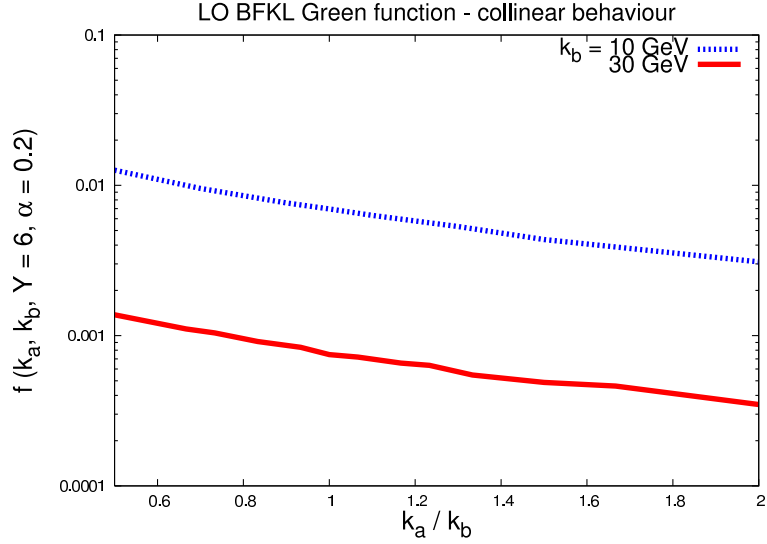}
    \caption{Collinear behaviour of the BFKL gluon Green function for different values of the reference scale $k_a$}
  \label{BFKLcollinear}
\end{figure}

\begin{figure}[htbp]
  \centering
  \includegraphics[width=10cm,angle=0]{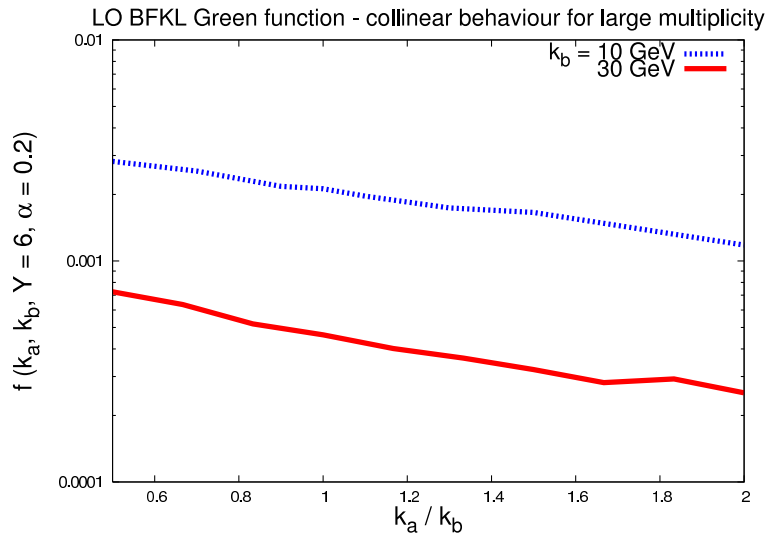}
    \caption{Collinear behaviour of the BFKL gluon Green function for different values of the reference scale $k_a$, 
    considering the contributions of configurations with more than 10 emissions.}
  \label{BFKLcollinearcut}
\end{figure}

The collinear/anticollinear limits, at leading order, are not as interesting as in the CCFM case since the Green function 
has the functional form $g(k_a / k_b) / (k_a k_b)$.  Changing the reference scale $k_b$ does not bring any new features as we can see 
in Fig.~\ref{BFKLcollinear}, where we chose $Y=6$. 
When imposing a cut to only keep high-multiplicity contributions we obtain a similar result, see Fig.~\ref{BFKLcollinearcut}

\begin{figure}[htbp]
  \centering
  \includegraphics[width=10cm]{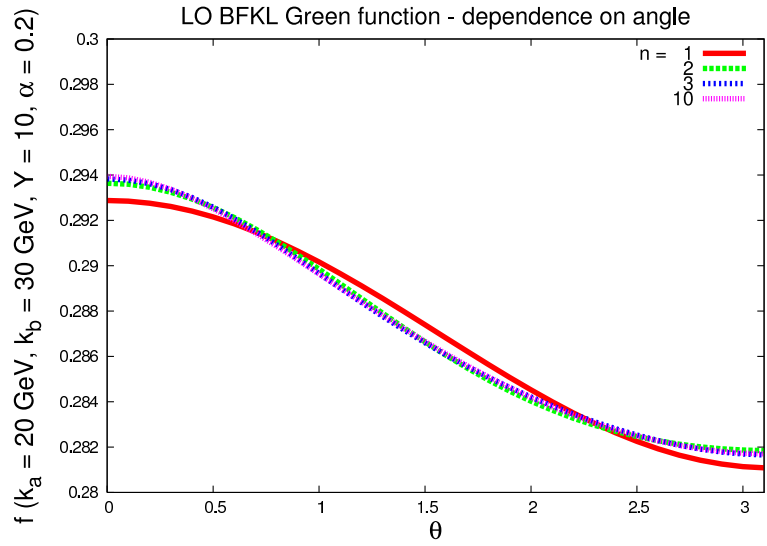}
    \caption{Azimuthal angle dependence of the BFKL Green function. We plot the result for the sum 
    in Eq.~(\ref{GGF1}) with up to $n = 1, 2, 3, 10$ terms.}
  \label{BFKLangles1}
\end{figure}

\begin{figure}[htbp]
  \centering
  \includegraphics[width=10cm,angle=0]{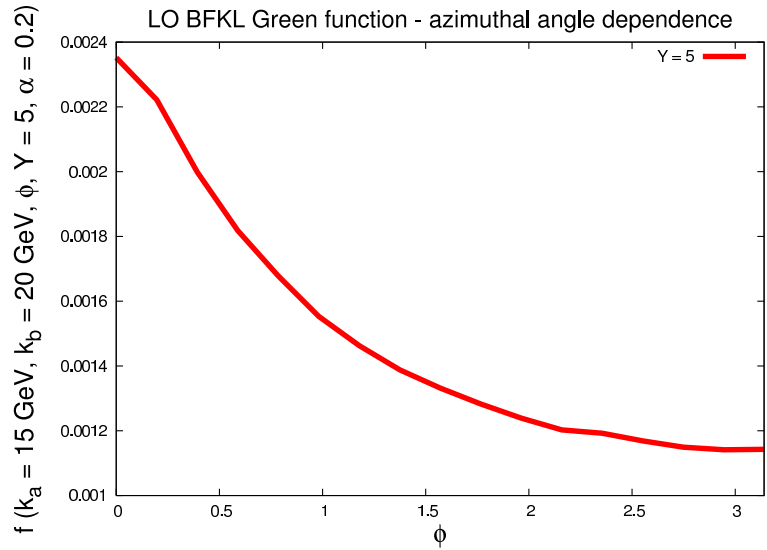}
    \caption{Full dependence on the azimuthal angle of the BFKL gluon Green function.}
  \label{BFKLangle}
\end{figure}

\begin{figure}[htbp]
  \centering
  \includegraphics[width=11cm]{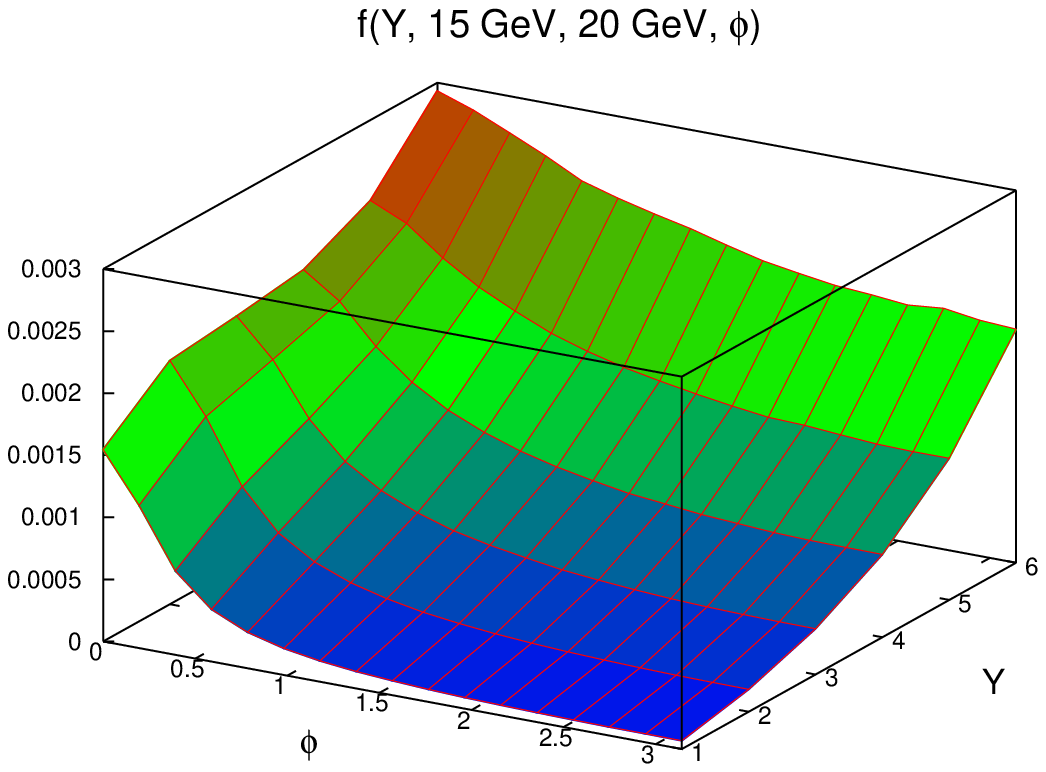}
  \caption{The combined dependence of the BFKL gluon Green function on the azimuthal angle and rapidity, for fixed values of the modulus of the boundary momenta.}
  \label{BFKLridge}
\end{figure}

\begin{figure}[htbp]
 \centering
  \includegraphics[width=11cm]{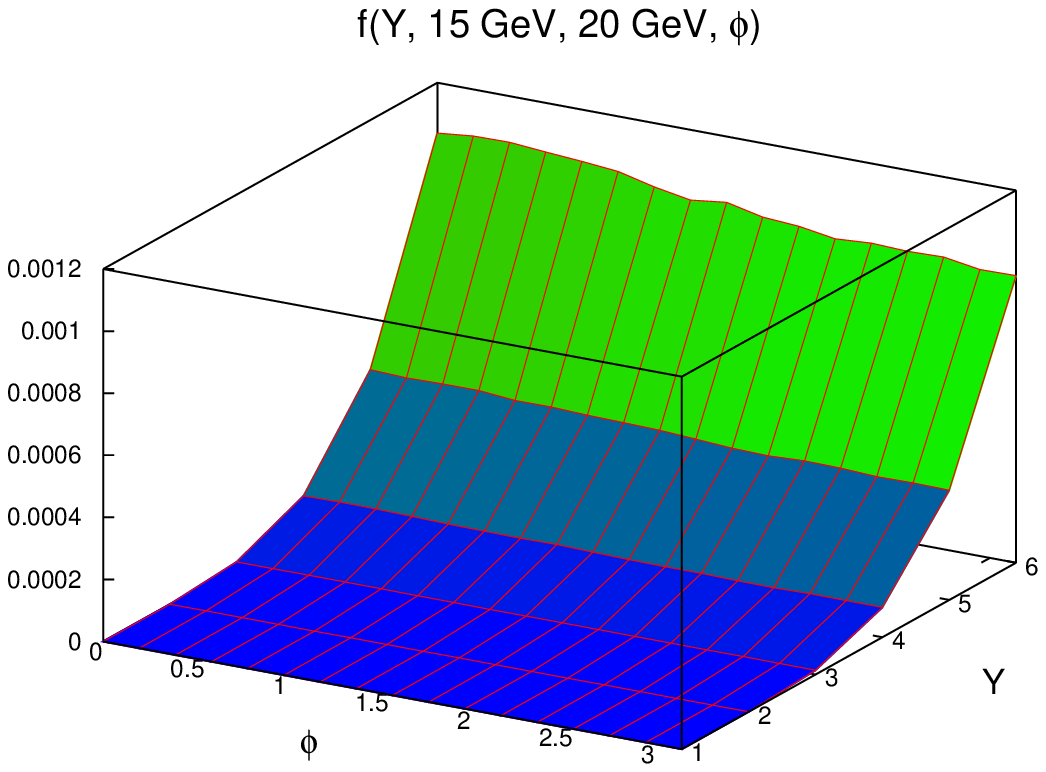}
  \caption{The combined dependence of the BFKL gluon Green function on the azimuthal angle and rapidity, for fixed values of the modulus of the boundary momenta, considering the contributions of configurations with more than 10 emissions.}
   \label{BFKLridgecut}
\end{figure}

Apart from the different diffusive behaviour, it is in the azimuthal angle dependence where we find more differences between the 
BFKL and the CCFM approaches. The expansion on Fourier components in the azimuthal angle can be written as
\begin{eqnarray}
f^{\rm BFKL} \left({\bf k}_a, {\bf k}_b,{\rm Y}\right) &=&
\sum_{n=-\infty}^{\infty} f^{\rm BFKL} _n\left(k_a, k_b, {\rm Y}\right) e^{i n \theta},
\label{GGF1}
\end{eqnarray}
with the coefficients being 
\begin{eqnarray}
f^{\rm BFKL} _n\left( k_a, k_b, {\rm Y}\right) &=& 
\frac{1}{\pi k_a k_b}  \int \frac{d \gamma}{2 \pi i} 
\left(\frac{k_a^2}{\vec{k}_b^2}\right)^{\gamma-\frac{1}{2}}
e^{\alpha \chi_n (\gamma) {\rm Y}}.
\end{eqnarray}
When performing our Monte Carlo analysis we obtain these coefficients using the momentum 
space numerical solution projecting on angles, i.e.
\begin{eqnarray}
f^{\rm BFKL} _n\left({k}_a, {k}_b, {\rm Y}\right) &=& 
\int_0^{2 \pi} \frac{d\theta}{2 \pi} \, f^{\rm BFKL}  \left({\bf k}_a, {\bf k}_b, {\rm Y}\right) 
\cos{\left(n ~ \theta\right)}.
\end{eqnarray}
In Fig.~\ref{BFKLangles1} we show how the convergence in $n$ for the Green function is very fast. 
In this example we can see that 10 terms in the series are enough to reach a very 
good approximation to the final solution. 
This agrees with the general behaviour obtained with our Monte Carlo analysis in Fig.~\ref{BFKLangle}.  
The combined plots in rapidity and azimuthal angle for full and high multiplicities are given in Fig.~\ref{BFKLridge} and 
Fig.~\ref{BFKLridgecut}. 

\begin{figure}[htbp]
  \centering
  \includegraphics[width=10cm,angle=0]{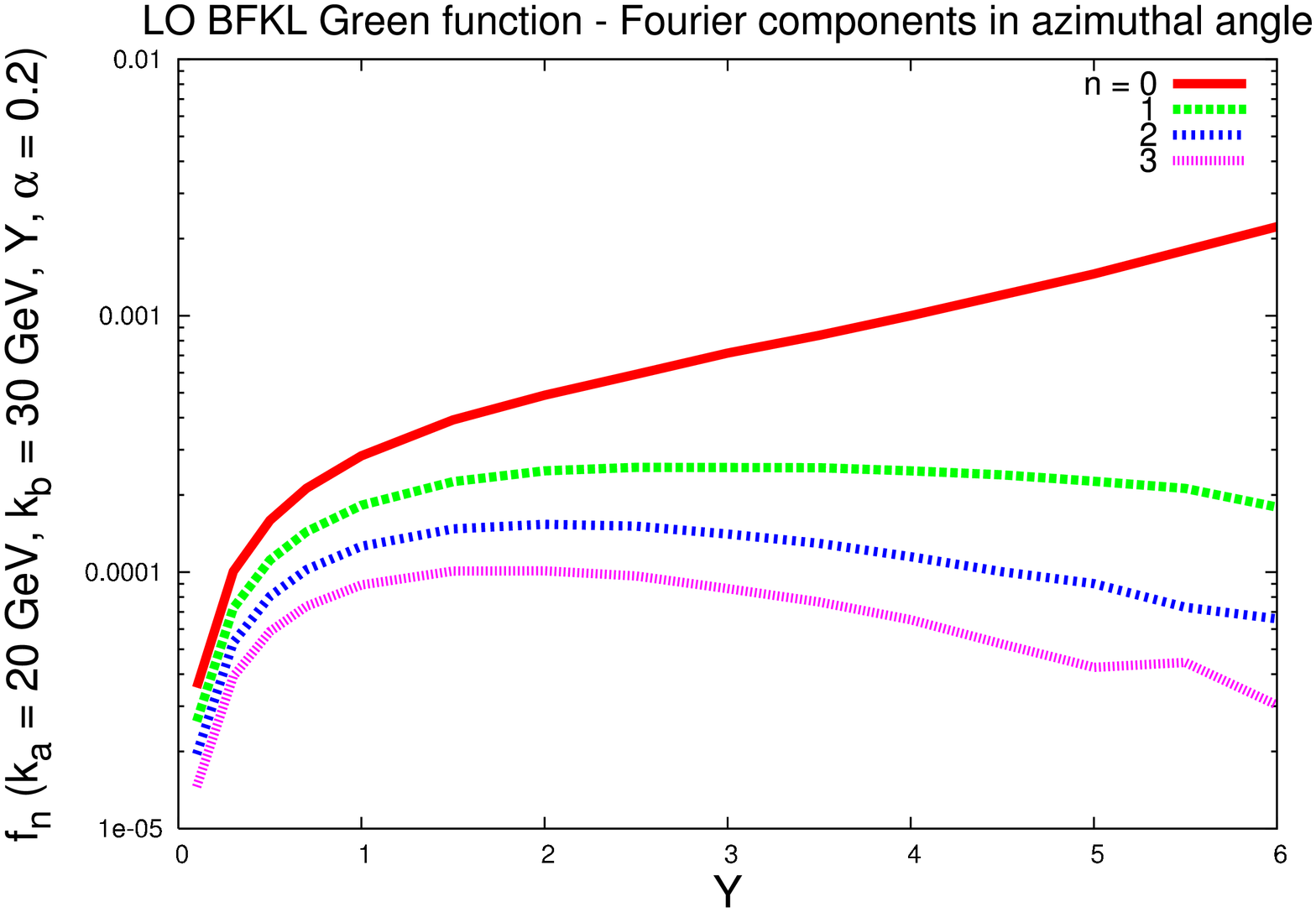}
    \caption{Variation with rapidity of the different components of the Fourier expansion on the azimuthal angle of the BFKL gluon Green function.}
  \label{BFKLfourier}
\end{figure}

Very importantly, in the BFKL approach the Fourier projections in $\theta$ have a very different behaviour to those in the CCFM formalism. As 
we can observe in Fig.~\ref{BFKLfourier} the $n=0$ projection also rises as in the CCFM case in Fig.~\ref{CCFMFourier}. 
However, the non-zero $n$ BFKL components do not rise with $Y$ while the CCFM ones do rise, see, again, 
Fig.~\ref{CCFMFourier}. This indicates that any observable sensitive to these higher angular components will have a completely different behaviour in both theories. There has been recent progress in the study of the azimuthal angle dependence within the BFKL approach, see, {\it e.g.}, Refs.~\cite{Kwiecinski:2001nh,Vera:2006un,Vera:2007kn,Vera:2008bz,Chachamis:2009ks,Baranov:2009zz,Marquet:2007xx,Colferai:2010wu}. In future publications we will present observables where 
this different azimuthal angle dependence might help discriminate between BFKL and CCFM contributions. A natural candidate is the 
azimuthal angle decorrelation between jets produced in the central region of rapidity and a jet emitted along the direction of one of the 
hadrons in the Large Hadron Collider. 

\section{Conclusions and scope}

In this letter we have compared two Monte Carlo implementations of the CCFM and BFKL formalisms for the 
description of small $x$ observables. The main difference between them from the theoretical point of view is the introduction 
of QCD coherence effects in the CCFM equation. We have found that the symmetric diffusion into infrared and ultraviolet 
regions of phase space characteristic of the BFKL parton evolution is broken in the CCFM case, where the infrared scales 
play a dominant role. As our main result we have found that the higher Fourier components in the gluon Green function 
have a very different behaviour in both theories, rising with energy in the CCFM case and decreasing in the BFKL one. 
It will be very interesting to trace these differences at an observable 
level~\cite{Deak:2008ky,Deak:2009xt,Deak:2010dg,Deak:2010gk} and to implement higher order 
corrections~\cite{Fadin:1998py,Ciafaloni:1998gs,Salam:1998tj,Vera:2005jt,Schmidt:1999mz,
Forshaw:1999xm,Chachamis:2004ab,Bartels:2006hg,Caporale:2007vs,Vera:2008bz} to evaluate their effects on them. These lines of research will be the subject of our future investigations. \\
\\
{\bf \large Acknowledgements}\\\\
This work has been partially supported by the European Comission under contract LHCPhenoNet (PITN-GA-2010-264564) and the Comunidad de Madrid through Proyecto HEPHACOS ESP-1473. M. D. acknowledges the support of the European Union through the Marie Curie Research 
Training Network  ``UniverseNet" (MRTCN-CT-2006-035863).

\end{document}